\documentclass[doublecol]{epl2} 
\usepackage{color}
\usepackage{soul}
\usepackage{mathtools}
\usepackage{amsmath}
\usepackage{graphics}
\usepackage{epsfig}
\usepackage{amssymb}
\usepackage{subfigure}
\usepackage{pifont}
\usepackage[T1]{fontenc}
\usepackage{textcomp}
\usepackage{amsmath}
\usepackage{amssymb}
\usepackage{wasysym}
\usepackage{graphicx}
\makeatletter
\usepackage{dcolumn}
\usepackage{bm}
\usepackage{float}
\usepackage{mathtools}
\usepackage{appendix}

\@ifundefined{textcolor}{}
{%
 \definecolor{BLACK}{gray}{0}
 \definecolor{WHITE}{gray}{1}
 \definecolor{RED}{rgb}{1,0,0}
 \definecolor{GREEN}{rgb}{0.3,.54,0}
 \definecolor{BLUE}{rgb}{0,0,1}
 \definecolor{CYAN}{cmyk}{1,0,0,0}
 \definecolor{MAGENTA}{cmyk}{0,1,0,0}
 \definecolor{YELLOW}{cmyk}{0,0,1,0}
 \definecolor{ORANGE}{rgb}{1,0.55,0}
 \definecolor{GRAY}{gray}{0.5}

 \definecolor{LCYAN}{rgb}{0,1,1}
 
 }

\newcommand{\cor}[1]{{#1}}
\newcommand{\corr}[1]{{#1}}

\title{Swimming bacteria in Poiseuille flow: the quest for active Bretherton-Jeffery trajectories}

\shorttitle{Swimming bacteria in flow} 

\author{Gaspard Junot$^1$ \and Nuris Figueroa-Morales$^2$ \and Thierry Darnige$^1$ \and Anke Lindner$^1$ \and Rodrigo Soto$^4$ \and Harold Auradou$^3$ \and Eric Cl\'ement$^1$ }
\shortauthor{G. Junot \etal}
\institute
{\inst{1} Laboratoire PMMH-ESPCI Paris, PSL Research University, \cor{Sorbonne Universit\'e and Denis Diderot, 7, quai Saint-Bernard}, Paris, France.\\
\inst{2} Department of Biomedical Engineering, The Pennsylvania State University, University Park, PA 16802, USA.\\
\inst{3}Laboratoire FAST, Univ. Paris-Sud, CNRS, Universit{\'e} Paris-Saclay,  F-91405, Orsay, France.\\
\inst{4} Physics Department, FCFM, Universidad de Chile, Av. Blanco Encalada 2008, Santiago, Chile.
}


\abstract{Using a 3D Lagrangian tracking technique, we determine experimentally the trajectories of non-tumbling {\it E. coli} mutants swimming in a Poiseuille flow. We identify a typology of trajectories in agreement with a kinematic "active Bretherton-Jeffery" model, featuring an axi-symmetric self-propelled ellipsoid. In particular, we recover the "swinging" and \cor{"shear tumbling"} kinematics predicted theoretically by Z{\"o}ttl \etal\cite{zottl2012nonlinear}. Moreover using this model, we derive analytically new features such as quasi-planar piece-wise trajectories, associated with the high aspect ratio of the bacteria, as well as the existence of a drift angle around which bacteria perform closed cyclic trajectories. However, the agreement between the model predictions and the experimental results remains local in time, due to the presence of Brownian rotational noise.}

\begin{document}

\maketitle

Understanding the motility and spreading of microorganisms in complex environments, sometimes undergoing significant flow variations, is relevant to many fundamental and technological issues. For instance, this is a crucial question in the context of medicine, as motility can control several physiological functions (e.g. spermatic transport in \corr{biological channels} \cite{bretherton1961rheotaxis, Denissenko2012sperm , riffell2007sex}, upstream contamination of urinary tracks \cite{uppaluri2012flow, Figueroa2015, Figueroa2019_Contamination} or virulence factors \cite{Josenhans2002}). It is also relevant to technologies of drug delivery \cite{nelson2010microrobots} and environmental studies aiming to understand the spreading of bio-contaminants in soils \cite{Ginn2002} or the building of ecological niches \cite{Bauer2018}. 

At the micro-hydrodynamical level, many models aiming to describe the transport processes of micro-swimmers in flow are based on a simple representation initiated by Jeffery\cite{Jeffery1922motion} and later completed by Bretherton \cite{bretherton1962motion}. The Bretherton--Jeffery (B-J) description assesses the changes of orientation of an axisymmetric ellipsoid in a Stokes flow, performing so called Jeffery orbits. The active version of the model (aka ``the active B-J model'') is completed with a swimming velocity added to the local flow velocity \cite{saintillan2013active}. In the active B-J model the ellipsoid orientation determines the swimming direction and the \cor{orientation} dynamics thus directly translate into complex particle trajectories. Passive particles in contrast do not cross streamlines and are merely transported downstream with the flow while tumbling \cor{with the velocity gradient}. 
In the presence of Brownian rotational noise, orientation distributions can be determined for passive particles in flow\corr{s} \cite{saintillan2013active,zottl2019dynamics,kim2017monitoring}. By including noise in the kinematic equations, the active B-J model is the base for many recent statistical-mechanics studies or hydrodynamic dispersion models \cite{saintillan2013active}. This is for example used to describe the emergence of a viscous response for active fluids \cite{saintillan2013active,Guzman2019} or the mean transport properties of bacteria in microfluidic channels\cite{bees2010dispersion,rusconi2014bacterial,chilukuri2014impact,creppy2019effect,chilukuri2015dispersion}. It can also serve to understand the motion of microorganisms in the presence of an external field such as magnetotactic bacteria moving in a flow in the presence of a magnetic field \cite{saintillan2018,vincenti2018}, an algae in a light intensity gradient \cite{garcia2013light} or in a gravity field for \corr{bottom-heavy} strains \cite{hill2002taylor}.

Interestingly, from an in-depth analysis of these kinematic equations without noise, Z{\"o}ttl \etal \cite{zottl2012nonlinear,zottl2013periodic} identified mathematical features and associated them with the emergence of cycloid trajectories. In some cases the kinematics can be mapped onto a dynamical Hamiltonian problem with conserved constants of motion\cite{zottl2012nonlinear,zottl2013periodic}. This model relies on a simplified vision of the bacterial shape and swimming behavior. For instance, the chirality of the bundle responsible of rheotactic effects \cite{Marcos2012, mathijssen2018oscillatory} or the flexibility of the flagella \cite{son2013bacteria,reigh2012synchronization,sporing2018hook} are neglected.

\cor{Despite the importance of this fundamental model, experimental validation and proof of its applicability to bacterial trajectories is still lacking. 
Tracking experiments of bacteria under flow have already been performed in 3D \cite{molaei2016succeed} but no B-J trajectories have been reported. A first experimental observation of these trajectories has been performed by Rusconi \etal \cite{rusconi2014bacterial} but only in the form of a 2D projection. 
Bacterial trajectories being \corr{3D} no full characterization of the dynamics and no direct comparison with the active B-J model could be performed.}

\begin{figure}[t!]
	\includegraphics[width=8.5cm]{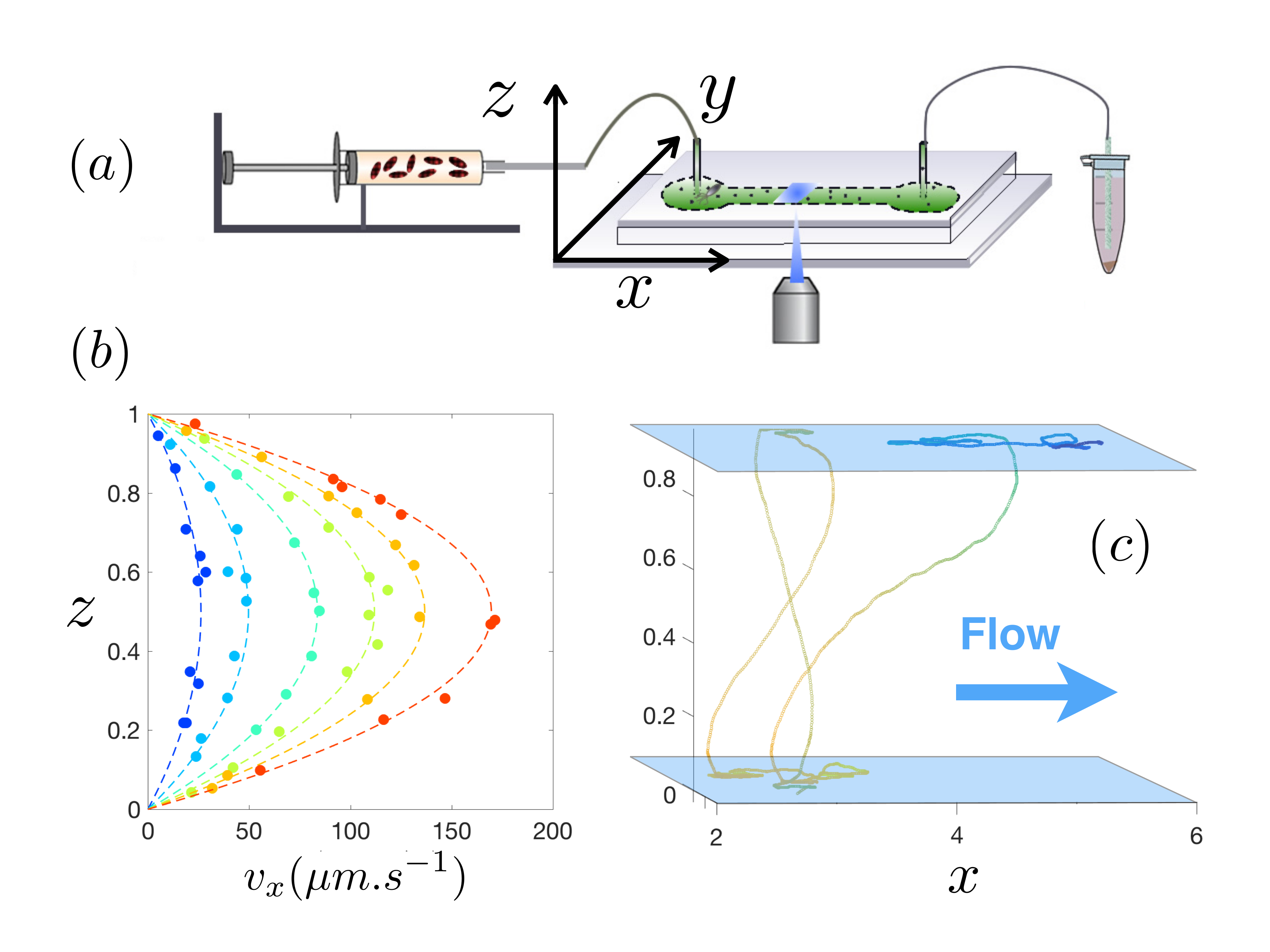}
	\vspace{0 mm}
	\caption{(a) Experimental set-up: a syringe pump injects a \corr{bacterial} suspension at constant flow rate into a channel of rectangular cross section \cor{(height $h=100~\mu m$ and width $W=600~\mu m$)}. (b) Velocity profiles, $v_x(z)$, obtained by tracking of tracer particles ($x$ is the flow direction). From blue to red the flow rates are $Q=1, 2, 3, 4, 5, 6~nL/s$. (c) Example of a 3D track of $\Delta$-CheY \textit{E. coli} mutant (smooth swimmer) obtained at $Q = 1~nL/s$. \cor{$x$, $y$ and $z$ have been made dimensionless using the channel height $h$ whereas $v_x$ is given in $\mu m/s$}.}
	\vspace{0 mm}
	\label{Set_up}
\end{figure}

In this letter, we investigate experimentally the 3D motion of {\it Escherichia coli} (\textit{E. coli}) bacteria in a \cor{plane Poiseuille flow}. We use a strain for which the tumbling process is inactivated and reorientation is due to the hydrodynamic shear and Brownian noise. Using a Lagrangian tracking device \cite{Darnige2016, figueroa20183d}, we identify the typology of many experimental 3D trajectories at different flow rates and compare them with the simulated trajectories stemming from the noiseless B-J model. Thanks to the 3D observation of the trajectories, we identify features such as swimming planes and drift angles, providing a better understanding of bacterial transport under flow. We derive several analytic solutions from the B-J model allowing a precise comparison with the experimental observations. Overall, we find good agreement between the experimental tracks and the theoretical predictions of the B-J model. The agreement seems robust despite the possibly more complex shape and swimming properties of real bacteria when compared to the simplified assumptions of an ellipsoid swimmer model. However,  this agreement remains local in time due to the cumulative influence of the  rotational noise on the trajectory.

\section{Set-up and protocol}
The measurements take place in a microfluidic channel of rectangular cross-section (height $h = 100 ~\mu m$, width $W = 600 ~\mu m$), made in Polydimethylsiloxane (PDMS) using standard soft-lithography techniques. Flow is imposed through the channel via a Nemesys syringe pump (dosing unit Low Pressure Syringe Pump neMESYS 290N and base Module BASE 120N). We set our region of interest in the center of the channel \cor{with respect to its width and consider only trajectories which are at least 100$\mu m$ away from the lateral walls.}

\begin{figure*}[t!]
	\includegraphics[width=18cm]{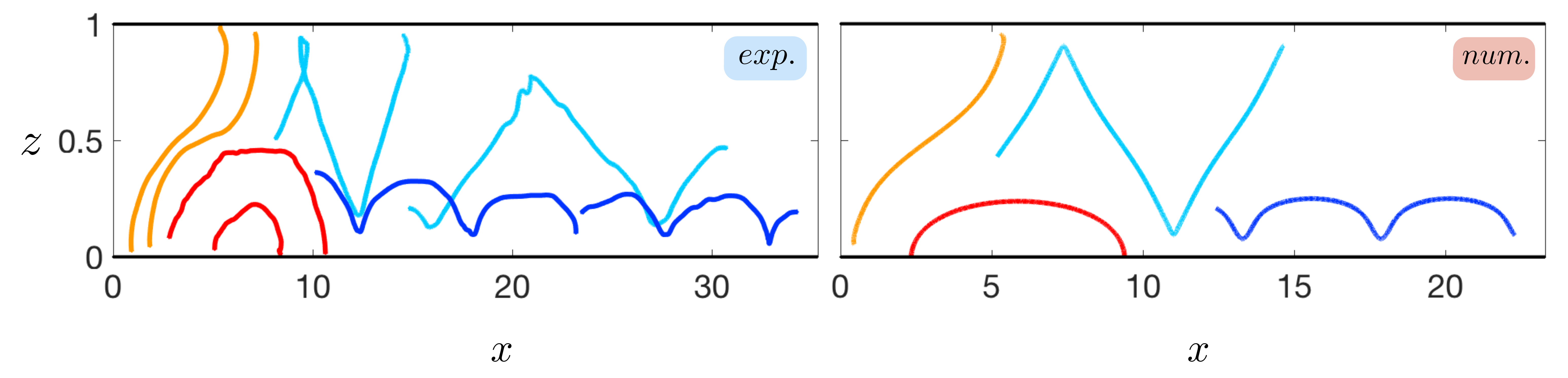}
	\vspace{-6 mm}	
	\caption{Typology of different \cor{3D} bacterial trajectories in Poiseuille flow \cor{projected} in the shear plane $z$-$x$ for different values of the parameters $A=V_b/4V_M$ and $\beta$ (a) Experimental trajectories: \cor{for each type of trajectory (different colors) two experimental trajectories are shown:} \cor{from left to right: A = 0.067, 0.062 (orange),  A = 0.15, 0.057 (red), A = 0.072, 0.053 (light blue) and A = 0.031, 0,020 (dark blue)}. (b) Numerical trajectories: from left to right: $A$ = 0.068, 0.15, 0.072, 0.031 and $\beta$ = 0.95, 0.95, 0.95, 0.80. 4 types of trajectories are observed:  (i) beginning and ending at the same wall (red), (ii) starting and ending at a different wall (orange), (iii) performing cycloid motion in a half channel (dark blue) and (iv) in the whole channel (light blue). For the numerical trajectories the values of $\beta$ have been chosen such as to reproduce qualitatively the experimental trajectories.}		
	\label{Typology}
	\vspace{0 mm}
\end{figure*}

The channel is visualized using a home-made Lagrangian tracking microscope \cite{Darnige2016} here used to track fluorescent swimming bacteria as well as passive tracers used for an accurate determination of the fluid velocity profile in the channel. By a visualization based feedback acting on a mechanical (horizontal) and piezoelectric (vertical) stage, the targeted object is kept \corr{close to} the center of the visualization field and \corr{in} focus \corr{on} an inverted microscope (Zeiss-Observer, Z1 with an objectif C-Apochromat 63x/1.2 W). Images of the tracked objects are acquired at $100$ Hz with a Hamamatsu Orca-flash 4.0 camera. Simultaneously, the three-dimensional positions of the object are recorded.
Figure~\ref{Set_up}(b) shows flow velocity profiles $v_x (z)$ obtained by tracking fluorescent beads of diameter 1$~\mu$m. Each dot is the mean velocity of a tracked bead, the dashed lines are parabolic fits used to obtain the maximal velocity at the center of the channel, $V_{M}$.

The bacteria used \corr{here} are smooth swimmer mutants of an \textit{E. coli} (strain CR20, $\Delta$-CheY) that almost never tumble and were transformed with a plasmid coding for a yellow fluorescent protein (YFP). Bacteria are grown overnight at 30$^\circ$C until \corr{the} early stationary phase.
The growth medium is then removed by centrifuging the culture and removing the supernatant. The bacteria are resuspended in a Motility Buffer (MB) below the very low concentration of $3\times10^7$ bacteria per mL, in order to visualize one bacterium at a time and to minimize the interactions between bacteria. The suspension is supplemented with L-serine at 0.08g/mL and polyvinyl pyrrolidone (PVP) at 0.005$\%$; L-serine maintains good motility for a few hours and PVP is used to prevent bacteria from sticking to surfaces. The solution is also mixed with Percol (1:1) to avoid \corr{bacterial} sedimentation. The experiments are performed at a temperature \corr{of} $25^{o}$C. 

Hundreds of trajectories are recorded at different flow rates $Q$, ranging from 1 to 6 $nL/s$, corresponding respectively to $V_{M}$ between $(28 \pm1.9)\corr{~\mu \text{m/s}}$ and $(168 \pm5.4)~\mu \text{m/s}$ and maximal shear rates $\dot{\gamma}_M=4 V_{M}/h$ between $(1.12 \pm0.076)~\corr{\text{s}^{-1}}$ and $(6.72\pm0.22)~\text{s}^{-1}$. 
Here we focus on bulk trajectories \cor{that are at least $5~\mu m$ away from the top and the bottom walls. Surface effects are \cor{known} to modify \corr{bacterial} trajectories close to channel walls \cite{molaei2014failed,mathijssen2018oscillatory} but are not the focus of the present paper. They also play a role in setting the initial conditions of bulk trajectories through bacteria take off from the surfaces.} The instantaneous swimming velocity is obtained for individual bacteria by removing the local flow velocity from the Lagrangian \corr{bacterial} velocity. For each track, this instantaneous swimming velocities follow a Gaussian distribution, and its mean gives the swimming velocity $V_b$. The average over all tracks is $\langle V_b\rangle = (25 \pm 5) \mu \text{m/s}$. \cor{The bacterium orientation is then obtain by dividing the instantaneous swimming velocity by its magnitude.}

%
\section{Typology of experimental trajectories}

Under flow, a significant number of \cor{bacterial tracks} were recorded and classified into 4 categories: (i) trajectories starting and ending at the same channel wall, (ii) trajectories starting from a wall and crossing the mid plane ($z/h = 1/2$) to reach the opposite wall, (iii) trajectories oscillating in a half channel below or above the mid plane and (iv) trajectories oscillating in the bulk and crossing the mid-plane \corr{repeatedly}. Figure \ref{Typology}(a) shows examples of such trajectories. The cycloid-like trajectories of type (iii) and (iv) correspond to the \cor{shear tumbling} and swinging trajectories predicted theoretically by Z{\"o}ttl \etal \cite{zottl2013periodic}. Note that trajectories starting and ending at channel walls are frequent in our experiments. \cor{As} smooth swimmer bacteria \cor{spend} a significant time at the solid boundaries, many trajectories recorded were indeed initiated at a channel wall.

\section{The active Bretherton--Jeffery model}

Under the B-J assumptions, the body of the microorganism is modeled as an ellipsoid of length $l$ and diameter $e$, swimming at a velocity $V_b\mathbf{p}$. The effective ellipsoid coordinates are its centroid position $\mathbf{r} = (x, y, z)$ with an orientation vector $\mathbf{p}= (\cos \theta , \sin \theta\cos \phi, \sin \theta\sin \phi )$~ (see fig.~\ref{Fig_coord_BJ_model}). 
The velocity of the ellipsoid is then the vectorial sum of the swimming velocity and the local flow velocity: $\mathbf{V}=V_b\mathbf{p}+\mathbf{v}$. The Bretherton--Jeffery derivation describes also how an axisymmetric ellipsoid of aspect ratio $r=l/e$ is reoriented in a Stokes flow with a strain-rate tensor $\overline{\overline{\mathbf{E}}}=(1/2)\left[\nabla\mathbf{v}+(\nabla\mathbf{v})^{T}\right]$ and a rotation rate tensor $\overline{\overline{\mathbf{\Omega}}}=(1/2)\left[\nabla\mathbf{v}-(\nabla\mathbf{v})^{T}\right]$. 
In the absence of rotational noise, the kinetic equation governing $\mathbf{p}$ is: $\mathbf{\dot{p}}=(\overline{\overline{\mathbf{I}}}-\mathbf{p}\mathbf{p})(\beta \overline{\overline{\mathbf{E}}}+\overline{\overline{\mathbf{\Omega}}})\mathbf{p} $, with $\overline{\overline{\mathbf{I}}}$ the identity tensor and  $\beta=(r^{2}-1)/(r^{2}+1)$  the Bretherton parameter. For a Poiseuille flow $\mathbf{v}= 4 V_{M} (1-\frac{z}{h})\frac{z}{h} \mathbf{e}_x $, the trajectories are controlled by the dimensionless parameters $A=V_b/4V_M$ fixing the ratio between the bacterium velocity and the maximal flow velocity and $\beta$. In the following, we rescale all distances by the channel height $h$ and time with $\dot\gamma_M$. The swimmer positions and orientations are then given by five coupled dynamical equations \cor{(consistent with the equations presented in \cite{zottl2013periodic})} implying three adimensionalized position coordinates $(x(t), y(t) , z(t))$

\begin{equation} \label{kinematics_positions}
\begin{gathered}
\begin{aligned}
\dot{x} &= A\cos(\theta) + z(1-z),  \\
\dot{y} &= A\sin(\theta)\cos(\phi),  \\
\dot{z} &= A\sin(\theta)\sin(\phi), 
\end{aligned}
\end{gathered}
\end{equation}
and two angular coordinates $\theta $ and $\phi$
\begin{equation}\label{kinematics_angles} 
\begin{gathered}
\begin{aligned}
 \dot{\theta } &= \frac{1}{2}\sin(\phi)[\beta\cos(2\theta) -1](1-2z),  \\
\sin(\theta) \dot{\phi }&= \frac{1}{2}(\beta -1)\cos (\phi)\cos(\theta).
\end{aligned}
\end{gathered}
\end{equation}

\begin{figure*}[t!]
\includegraphics[width=18cm]{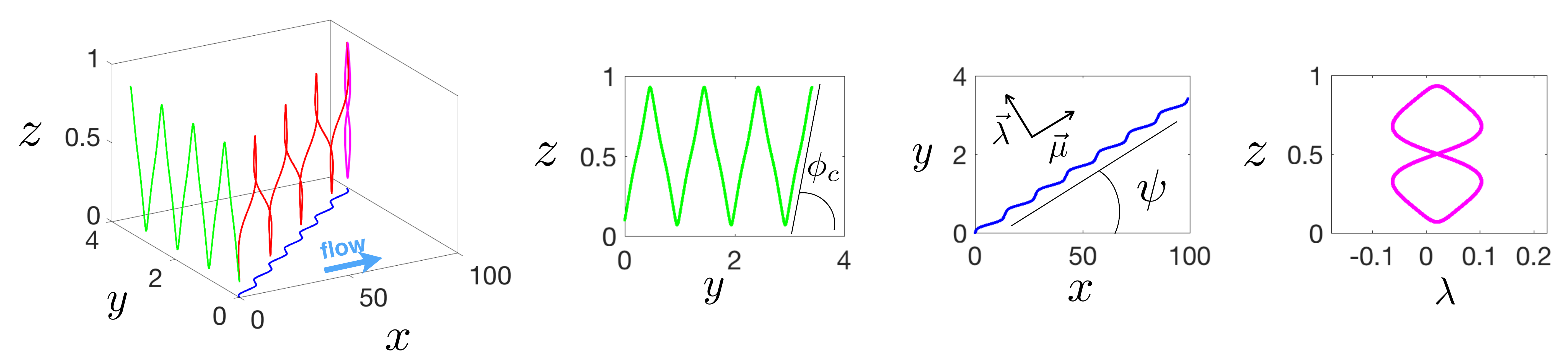} 
\vspace{-6 mm} 
\caption{Example of a type (iv) \cor{swinging trajectory}(see fig.~\ref{Typology}) computed using $A = 0.04$ and $\beta = 0.95$. In the $y$-$z$ plane (green) the bacterium swims with two nearly constant angles $\phi_c$ or $2\pi-\phi_c$. In the $x$-$y$ plane (blue), one observes the cycloid-like trajectory with a drift angle $\psi$. By rotating the trajectory by the drift angle $\psi$ \cor{(computed using the analytical expression given by eq.~\eqref{TanPsi})} around the $z$-axis, the trajectory collapses in the $\lambda$-$z$ plane into a closed orbit (in pink).}
\label{Fig_numeric_projection}
\end{figure*}

\section{Model predictions and comparison to experimental observations} 

Using the active B-J model without noise, we first numerically determine typical bacterial trajectories. We then derive closed form expressions for the phase portraits $z(\theta)$ and $\phi(\theta)$ as well as an analytical expression for the \corr{bacterial} drift angle  $\psi$ with respect to the flow direction. We call $z_0$, $\phi_0$ and $\theta_0$ the set of initial conditions. The model predictions are then compared to experimental observations. \cor{Note that surface effects are neglected in our model and quantitative comparisons will only be performed on bacterial trajectories that are at least $5~\mu m$ away from the top and the bottom walls.}

\subsection{Trajectories}
As no closed form solutions are available for the B-J model we solve the model numerically. The numerical trajectories displayed on figs.~\ref{Typology}, \ref{Fig_numeric_projection} and \ref{Fig_comparaison}, are obtained by numerical integration of eqs.~(\ref{kinematics_positions}) and (\ref{kinematics_angles}) simply using an Euler scheme. Typical experimental observations are reproduced by the numerical trajectories as can be seen from fig. \ref{Typology} where all trajectory types observed are displayed. The detailed properties of these trajectories are discussed below.

\begin{figure}[h!]
	\includegraphics[width=8.5cm]{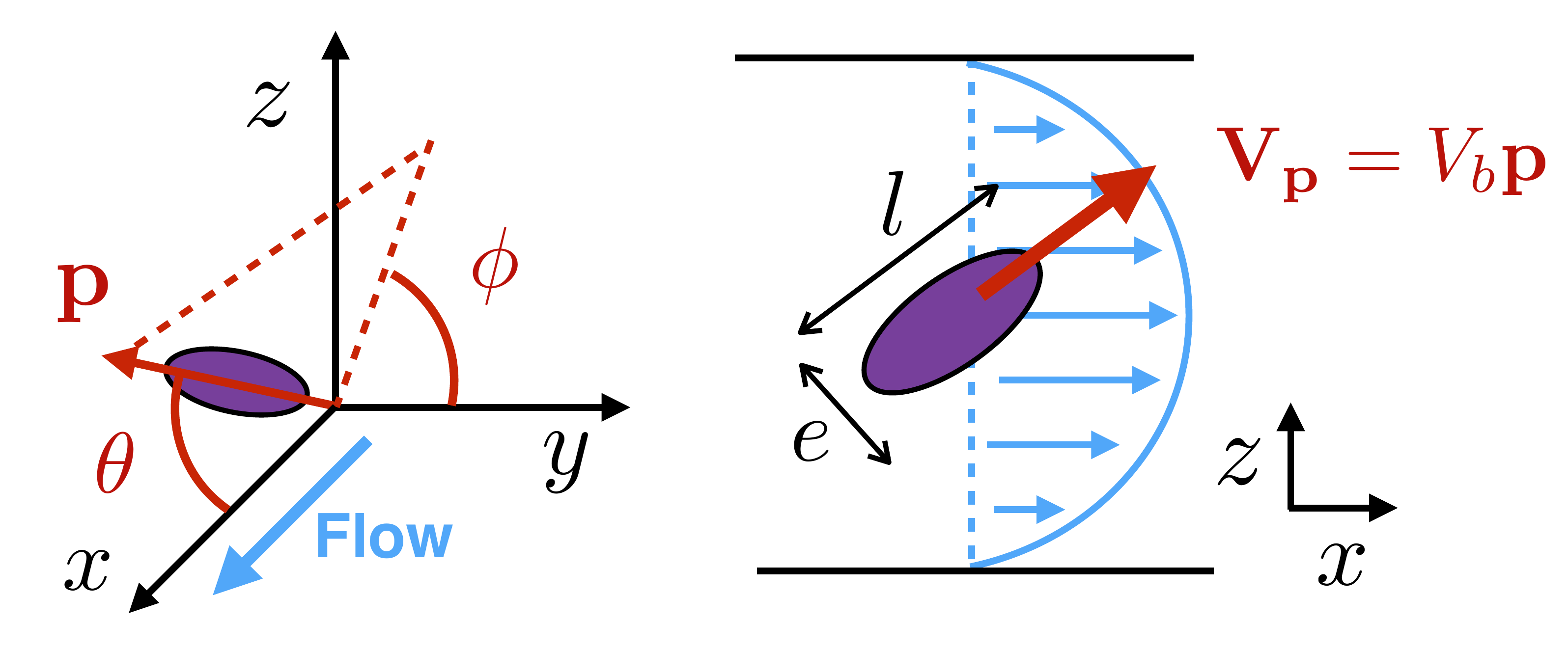}
	\vspace{0 mm} 
	\caption{Parametrization of an effective ellipsoid swimming in a Poiseuille flow.} 
	\label{Fig_coord_BJ_model}
\end{figure}

\subsection{Swimming in planes of nearly constant angles}

\begin{figure*}[t!]
	\centering 
        \includegraphics[width=13.5cm]{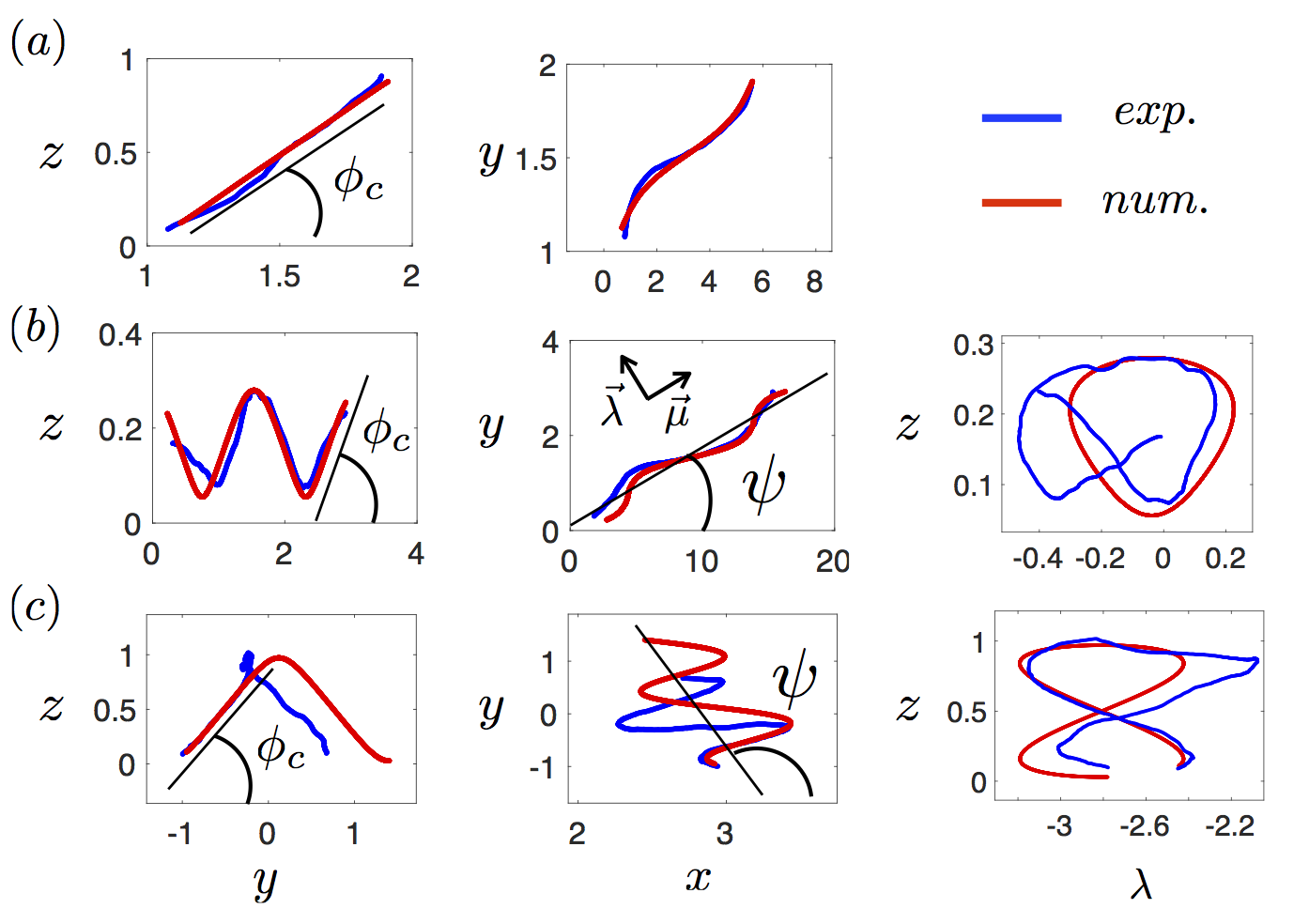} 
	\vspace{0 mm} 
	\caption{Projection in the y-z, x-y and z-$\lambda$ planes of experimental (blue lines) and numerical trajectories (red lines).
		The numerical trajectories are computed using the parameter $A$ determined experimentally and using as initial conditions the positions and angles of the bacterium in the middle of the track considered. The only fitting parameter is $\beta$. The projection in the $\lambda$-$z$ plane is performed using the value of $\psi$ computed with eq.~\eqref{TanPsi}. (a) type (ii) trajectory ($\beta = 0.90$ and $A = 0.059$). (b) type (iii) swinging trajectory ($\beta = 0.95$ and $A = 0.045$). (c) type (iv) shear tumbling trajectory ($\beta = 0.86$ and $A = 0.19$).} 
		\label{Fig_comparaison} 
\end{figure*}

An important remark can be made  from the angle variations $\dot{\phi }$ in eq.~(\ref{kinematics_angles}). 
For strongly elongated particles, i.e. $\beta \rightarrow 1$, the angle derivative is almost zero unless the angle $\theta$ reaches values \corr{of} $0$ or $\pi$. This property actually corresponds to a swimmer motion dwelling very close to a plane of constant angle $\phi_c$ until a flip occurs for particle orientations close to $\theta=0$ or $\pi$ in order to set the motion into the mirror plane defined by $\phi=2\pi-\phi_c$.
In the $y-z$ plane perpendicular to the flow direction $x$, these planes appear as lines of directions $\phi_c$ and ($2\pi-\phi_c$). The $\phi_c$ values are of course fixed by the initial conditions. 
These fixed planes hosting the bacterial motion, can be seen as a consequence of the absence of shear in \corr{the y direction}. In fig.~\ref{Fig_numeric_projection}, we display a numerical trajectory of type (iv) performing swinging motion, as well as its projection onto various planes. The planes of constant angle $\phi$ are clearly illustrated by the $y$-$z$ projection (in green). Figure \ref{Fig_comparaison} shows different experimental and numerical trajectories (type (ii), (iii) and (iv)) projected into the same planes as on fig.~\ref{Fig_numeric_projection}. From the projections \corr{into} $y-z$  (first column), one can clearly observe a tendency for a bacterium to swim in planes of nearly constant angle $\phi$ and also the presence of subsequent flipping between mirror planes. Hence this good agreement between the experimental \corr{observations} and the simulations indicates that the smooth \textit{E.coli} swimmers are well modeled by a $\beta$ value close to 1,  \cor{corresponding to very elongated objects}.

\subsection{Drift angle $\psi$}

For the numerical trajectories of type (iii) and (iv), i.e. for bacteria traveling in the bulk without touching the walls, another feature can be noticed. When projected into the $x$-$y$ plane (see for example fig.~\ref{Fig_numeric_projection}), these trajectories oscillate around a mean direction different from the flow direction hence defining a ``drift angle'' $\psi$. \cor{This drift can be seen} as the ratio of two displacements, one along or against the flow (the latest resulting from bacteria swimming upstream) and an a transverse displacement solely due to the bacterial activity.
The corresponding trajectories are periodic in $z$ (see fig.~\ref{Typology}) and correspond to closed trajectories in the ($z$, $\theta$, $\phi$) phase-space with a time periodicity T. \cor{Due to the bacterium activity and the dependence on $z$ of the local shear, this period T will be different from the period of the classical Jeffery orbit. Starting from a point in the ($z$, $\theta$, $\phi$) phase space, we define the period T as the time to go back to this point.} Since $\dot{x}$ and $\dot{y}$ only depend on $z$, $\theta$ and $\phi$ (all periodic functions), one can then define the displacements over one period T, along the flow, $\Delta x$, and perpendicular to the flow, $\Delta y$. The expressions of these two displacements are obtained by direct integration of eq.~(\ref{kinematics_positions}) and their ratio yields an expression for the tangent of the drift angle $\psi$. In Supplementary Material (Supplementarymaterial.pdf (SM)), we detail a closed-form derivation for the displacements $\Delta x$ and $\Delta y$ and show how an analytical expression for $\tan \psi $ can be obtained. Then the tangent of the drift angle:

\begin{equation}\label{TanPsi}
\tan \psi = f(A,\beta,z_{0},\theta_{0},\phi_{0}),
\end{equation}
is parametrized by the dimensionless numbers of the problem, $ A,\beta $ and the initial trajectory conditions $z_{0},\theta_{0},\phi_{0}$.
From the projection of the numerical trajectory of type (iv) into the shear plane $x$-$y$ in fig.~\ref{Fig_numeric_projection} the drift angle
$\psi$ is clearly identified. We define the $\mu,\lambda$-coordinates as shown in fig.~\ref{Fig_numeric_projection} respectively along and  perpendicular to the drift direction. We then obtain a remarkable property by projecting the trajectory in the plane $\lambda$-$z$ resulting from a rotation around $z$ by the angle $\psi$ obtained from the analytical expression (eq.~\ref{TanPsi}): \cor{each 3D B-J trajectory collapses onto a closed orbit in \corr{the} $z$-$\lambda$ plane \corr{with a shape depending} on the initial conditions of the trajectory and on the parameters A and $\beta$.} Similar results are observed for the experimental trajectories of type (iii) and (iv) shown together with corresponding numerical predictions on figs.~\ref{Fig_comparaison}(b) and (c). For all these cases the drift angle $\psi$ and the closed orbits are clearly visible. 
Note importantly, that the \corr{dependence} of the drift angle $\psi$ on  initial conditions $z_{0}, \theta_{0}$ and $\phi_{0}$ may lead to important consequences for the macroscopic  transport properties. For example, our calculations show that the direction of the drift is explicitly dependent on the sign of the angle $\phi_0$, which might be selected during the phase of detachment from solid boundaries through non trivial interaction processes between bacteria and the wall \cite{Frymier1995, schaar2015detention, bianchi2017holographic, mathijssen2018oscillatory}. Any biased distribution of initial orientations stemming from the boundary conditions will contribute to a net bacterial drift which could add up to the rheotactic contribution due to chirality as proposed by Marcos \etal \cite{Marcos2012}.

\subsection{Phase portraits}

The derivation of the phase portraits has previously been performed by Z{\"o}ttl \etal \cite{zottl2013periodic}. Here we chose a different angle parametrization (fig.~\ref{Fig_coord_BJ_model}) more suited to highlight the geometrical features we experimentally observe and thus rederive these results. We first calculate the phase portrait  $z(\theta)$. The ratio between $\dot{z}$ and $\dot{\theta}$ (eqs~\eqref{kinematics_positions} and \eqref{kinematics_angles}) yields $\frac{dz}{d\theta} = \frac{2A\sin(\theta)}{[\beta\cos(2\theta) -1](1-2z)}$, \corr{and} can be integrated to obtain the relation:

\begin{equation}\label{z_theta}
z_{\pm} = \frac{1\pm\sqrt{1+4B(\cos\theta)}}{2}
\end{equation}
\begin{equation}\label{z_thetaB}
B(\cos\theta) = -z_0(1-z_0) + \frac{A}{2a\beta}\ln\left[ \frac{(a+\cos\theta)(a-\cos\theta_0)}{(a-\cos\theta)(a+\cos\theta_0)}\right]
\end{equation}

where $a = \sqrt{\frac{\beta+1}{2\beta}}$.
The solutions $z_+$ and $z_-$ correspond respectively to sections of trajectories in the upper half $(0.5<z<1)$ or in the lower half $(0<z<0.5)$ of the channel. We then evaluate the phase portrait $\phi(\theta)$. By dividing $\dot{\phi }$ by $\dot{\theta}$ (eqs.~\eqref{kinematics_positions} and \eqref{kinematics_angles}) one obtains $\sin\theta\frac{d\phi}{d\theta} = \frac{(\beta -1)\cos (\phi)\cos(\theta)}{\sin\phi[\beta\cos(2\theta) -1]}$, yielding after integration:
\begin{equation}\label{cosphi_tantheta}
\begin{gathered}
\begin{aligned}
\Big| \frac{\cos\phi}{\cos\phi_0} \Big|= \Big| \frac{\tan\theta_0}{\tan\theta} \Big| \sqrt{\frac{1+r^2\tan^2\theta}{1+r^2\tan^2\theta_0}}.
\end{aligned}
\end{gathered}
\end{equation}

Figure \ref{Fig_noise}(a) shows good agreement of an experimental trajectory and the numerical prediction represented in the phase portraits. 
In addition, from the phase portraits displayed in figs.~\ref{Fig_noise}~(a-c), we can rationalize the prominence of trajectories staying close to a plane $\phi = \phi_c$ as long as the angle $\theta$ is not close to $0$ or $\pi$ (i.e. $\sin\theta$ going to zero). Then, in this last case, the model predicts a shift from the initial plane to \cor{the mirror plane} (see fig.~\ref{Fig_numeric_projection}). As already noticed this is indeed a robust feature observed experimentally for many trajectories. 
\section{Influence of  orientational noise}
For all trajectories observed experimentally, the features revealed by the active 3D B-J model have been recovered semi-quantitatively. However, even for the smooth swimmer strain used here, a quantitative agreement between simulation and experiments is only local in time. Indeed, after a relatively short observation period experimental trajectories deviate systematically from the numerical predictions, even for bacterial trajectories that remain far from the channel wall\corr{s}. We attribute this deviation to the presence of orientational noise in the experiment. Such erratic changes in bacteria orientation can be due to several reasons such as a mechanical bending of the flagellar bundle under shear\cite{tournus2015flexibility}, remnant tumbling processes or thermal fluctuations.
By the nature of the equations of motion \eqref{kinematics_positions} and \eqref{kinematics_angles}, any variation in the orientation produces cumulative large deviations in the positions, hence limiting the possibility to obtain global agreement \corr{on} the trajectories.
 
\begin{figure}[t!]
\includegraphics[width=8.5cm]{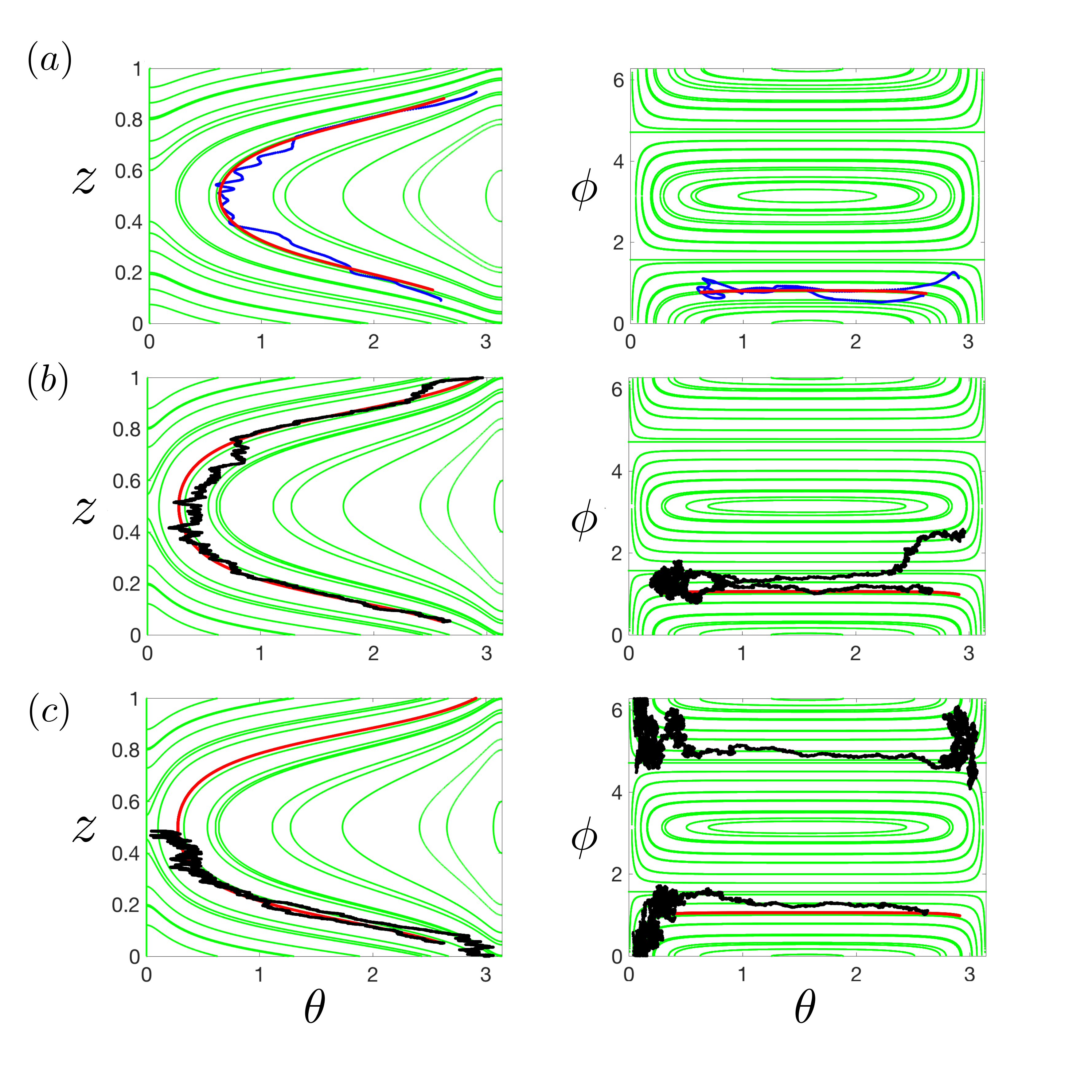}
\vspace{0 mm}
\caption{Phase portraits. The phase lines in green are obtained using eqs.~\eqref{z_theta} and \eqref{cosphi_tantheta}, blue line: experimental data, red and black lines are resp. numerical simulations without and with rotational noise projected into the phase space. (a) Phase portraits ($A = 0.059$ and $\beta = 0.90$). (b) and (c) phase portraits ($A=0.0625$ and $\beta = 0.97$). The trajectories in black show two different realizations of a numerical trajectory with noise ($D_r=1/47 s^{-1}$) simulated using the same initial conditions and parameters as for the red trajectory without noise. In panel (b) simulations with and without noise remain very close, whereas in panel (c) an important difference is observed.}
\vspace{0 mm}
\label{Fig_noise}
\end{figure}

To illustrate the influence of noise we display on fig.~\ref{Fig_noise} the $\theta$-$z$ and $\theta$-$\phi$ phase-spaces (green lines), parametrized by $A$ and $\beta$ corresponding to typical experimental realizations. To the B-J trajectories, we added an orientational noise term of amplitude $D_r = 1/47 s^{-1}$ (in black) \cite{figueroa20183d}  corresponding to the rotational diffusion of an ellipsoid of size $l = 8 \mu m $ and width $e = 1 \mu m$ (equations given in the SM).
Figures \ref{Fig_noise}(b) and (c) show the same phase portrait including a chosen numerical trajectory without noise (in red). Two different realizations with noise are shown in figs.~\ref{Fig_noise}(b) and (c) respectively, demonstrating that the presence of Brownian rotational noise can \corr{lead to very different trajectories for identical initial conditions.} 

\section{Discussion and conclusion} 
We have shown that the active B-J model reproduces semi-quantitatively the observed experimental trajectories of non tumbling \textit{E. coli} bacteria swimming in a Poiseuille flow.  In particular, we \corr{have proven} experimentally the existence of families of cycloid-like swinging and \cor{shear tumbling} trajectories as predicted by  Z{\"o}ttl \etal \cite{zottl2013periodic}. Therefore, in spite of the geometrical complexity of {\it E. coli} bacteria, the core of the Bretherton model associating a swimming bacterium \corr{with} an effective ellipsoid is validated experimentally.

We have shown the propensity to swim in planes of nearly constant angle $\phi=\phi_c$ along the flow and to repeatedly switch between \cor{$\phi=\phi_c$ and its mirror planes $\phi=2\pi-\phi_c$}, a robust feature recovered experimentally. We have established that this feature is associated with the long aspect ratio of the bacteria (Bretherton parameter $\beta \rightarrow 1$).  We \corr{have shown} that cycloid-like trajectories display a drift angle with the flow direction and we \corr{have proven} that after a rotation around the vertical axis, these oscillating B-J trajectories do collapse onto closed orbits. These properties are also recovered experimentally and we \corr{have provided} an analytical expression of this drift angle which will be crucial in the dispersion mechanism along the direction perpendicular to the flow. The drift angle strongly depends on the initial conditions and a bias in the initial orientation as for example induced by interactions with surfaces might lead to bacteria drift into specific directions. 

Crucial questions remain concerning the reorientation of the swimming angles due to rotational noise, which contributes to the hydrodynamic dispersion process (in the real $x$-$y$-$z$ space). Here we have shown that the \cor{randomization} process observed in the phase space is consistent, at least in magnitude, with a rotational Brownian motion, for an effective ellipsoid. It is, however, possible that other sources of randomization come into play such as the bundle flexibility \cite{Potomkin2017} partial debundling due to shear, \cor{wiggling effects \cite{hyon2012wiggling} or reorientation due to the chirality of the bacteria flagella \cite{Marcos2012}.}

\acknowledgments
The authors thank Dr Reinaldo Garc\'ia Garc\'ia for useful discussions. This work was supported by the ANR grant ``BacFlow'' ANR-15-CE30-0013 and the Franco-Chilean EcosSud Collaborative Program C16E03. N.F.M. thanks the Pierre-Gilles de Gennes Foundation for financial support. A.L. and N.F.M. acknowledge support from the ERC Consolidator Grant PaDyFlow under grant agreement 682367. RS acknowledges support of the Fondecyt Grant No. 1180791 and the Millenium Nucleus Physics of Active Matter of the Millenium Scientific Initiative of the Ministry of Economy, Development and Tourism (Chile).


\section{Supplementary material}

\subsection{Derivation of the drift angle $\psi$}

We recall here some expressions from the main document needed to perform the calculation of $\tan\psi$.

\begin{equation} \label{kinematics_positions}
\begin{gathered}
\begin{aligned}
\dot{x} &= A\cos(\theta) + z(1-z)  \\
\dot{y} &= A\sin(\theta)\cos(\phi)  \\
\dot{z} &= A\sin(\theta)\sin(\phi)   
\end{aligned}
\end{gathered}
\end{equation}

\begin{equation}\label{kinematics_angles} 
\begin{gathered}
\begin{aligned}
 \dot{\theta } &= \frac{1}{2}\sin(\phi)(\beta\cos(2\theta) -1)(1-2z)  \\
\sin(\theta) \dot{\phi }&= \frac{1}{2}(\beta -1)\cos (\phi)\cos(\theta) 
\end{aligned}
\end{gathered}
\end{equation}

\begin{equation}\label{z_theta}
z_{\pm} = \frac{1\pm\sqrt{1+4B(\cos\theta)}}{2}
\end{equation}
with:
\begin{equation}\label{z_thetaB}
 B(\cos\theta) = -z_0(1-z_0) + \frac{A}{2a\beta}\ln\left[ \frac{(a+\cos\theta)(a-\cos\theta_0)}{(a-\cos\theta)(a+\cos\theta_0)}\right]
\end{equation}

\begin{equation}\label{cosphi_tantheta}
\begin{gathered}
\begin{aligned}
 \Big| \frac{\cos\phi}{\cos\phi_0} \Big|= \Big| \frac{\tan\theta_0}{\tan\theta} \Big| \sqrt{\frac{1+r^2\tan^2\theta}{1+r^2\tan^2\theta_0}}
\end{aligned}
\end{gathered}
\end{equation}

From eqs.~(\ref{kinematics_positions}), we see that an extremum is reached when $\dot{z}=0$ i.e $\sin\phi=0$ or equivalently $\cos\phi^2=1$. Calling $\theta^*$ the angle corresponding to the $z$ extremum, and \cor{setting $\cos\phi^2$ to 1} in eq.~(\ref{cosphi_tantheta}) one obtains:

\begin{equation}\label{cos_thetastar}
\begin{gathered}
\begin{aligned}
\cos \theta^*_{\pm} =\pm \sqrt{ \frac{1+r^2\tan^2\theta_0\sin^2\phi_0}{1+\tan^2\theta_0(r^2\sin^2\phi_0+\cos^2\phi_0)}}
\end{aligned}
\end{gathered}
\end{equation}\\

From direct integration of eq.~\ref{kinematics_positions} for $x$ and $y$ we obtain:

\begin{equation}\label{deltaxy0}
\begin{gathered}
\begin{aligned}
\Delta x &= \int_{t_0}^{t_0+T}[ A\cos\theta+z(1-z)]\, dt\\
\Delta y &= \int_{t_0}^{t_0+T} A\sin\theta\cos\phi\, dt
\end{aligned}
\end{gathered}
\end{equation}

For a given trajectory, due to the periodicity, the values of $ \Delta x$ and  $ \Delta y$ do not depend on the specific choice of $t_0$, the starting time. Therefore, one can define the angle $\psi$ such that $\tan\psi = \Delta x /\Delta y $, as a quantity independent of the choice of initial position on the curve. To perform an explicit calculation \cor{of} $\tan \psi$, we choose to start from one of the extrema of $z$.

As we do not know the expression of the period T nor any expression of $z, \theta, \phi$ as function of time, we perform a change of variable $t$ into $\eta = \cos \theta$ in the integrals of eqs.~(\ref{deltaxy0}). Starting from a point in the phase space $z, \theta, \phi$, the period T is defined  as the time to go back to the same point. So after the change of variable the integral\cor{s} run from $\cos\theta(t_0)$ to  $\cos\theta(t_0+T)=\cos\theta(t_0)$. These integrals are not equal to zero because the change of variable is not bijective. We thus have to modify the integrand to \cor{ensure} a one-to-one correspondence as it is \cor{explained} in the following.   

The relation $\eta(\theta)$ is bijective in the interval $[0,\pi]$. However, as the function $t \rightarrow \theta(t)$ is not bijective on $[\theta(t),\theta(t+T)]$, the integral has to be split over domains to \cor{ensure} a one-to-one correspondence. Using eqs.~\ref{cosphi_tantheta} and \ref{cos_thetastar} one obtains $\sin^2\phi$ as a function of $\theta$ and $\theta^*$:


\begin{equation}\label{sin2phi}
\begin{gathered}
\begin{aligned}
\sin^2\phi &= \frac{\cos^2\theta^*-\cos^2\theta}{(1-\cos^2\theta)(a^2-\cos^2\theta^*)(r^2-1)}
\end{aligned}
\end{gathered}
\end{equation}
We also need an expression for the angle $\theta_{1/2} = \theta(z=0.5)$. Using eq.~(\ref{z_theta}) for $z=1/2$, we obtain:
\begin{equation}\label{costheta1_2}
\begin{gathered}
\begin{aligned}
\cos\theta_{1/2} &= \frac{a[\cos\theta_0(1+C_{1/2} )+a(1-C_{1/2} )]}{\cos\theta_0(1-C_{1/2}) +a(1+C_{1/2} )}\\
C_{1/2} &= \exp\left[{ \frac{ z_0(1-z_0)-\frac{1}{4}}{A}\sqrt{2(\beta+1)\beta} }\right]
\end{aligned}
\end{gathered}
\end{equation}
To express $d\theta$ as function of $\theta$ and $dt$, one replaces $z$ eq.~(\ref{z_theta}) and $\sin \phi$ eq.~(\ref{sin2phi}) in the expression of $\dot{\theta}$ in eq.~(\ref{kinematics_angles}). Then:
\begin{equation}
d\theta = - S \frac{1}{2}\sqrt{\sin^2\phi}(1-\beta\cos(2\theta))\sqrt{1+4B(\theta)}dt
\end{equation}
where $S$ is a sign function taking $\pm 1$ values such as to keep $dt>0$ on the domains of integration according to $\theta$ variations. Therefore, for the $\eta(t)$ variation, one obtains the relation:

\begin{equation}
\begin{gathered}
d\eta = S \beta |\sin \phi| \sqrt{1-\eta^2}(a^2-\eta^2)\sqrt{1+4B(\eta)}dt\\
\end{gathered}
\end{equation}
Therefore, to keep $t(\eta)$ bijective piece-wise  for the integrals in eq.~(\ref{deltaxy0}), the domains of integration are chosen between the extrema of $\theta(t)$. Eq. (\ref{kinematics_angles}) shows that $d\theta$ changes sign when $\sin \phi = 0$ or when the trajectory crosses the plane $z=1/2$. We perform an integration from the value $\eta^*=\cos(\theta^*_{-})$, which means we \cor{have to} start from $z_0 = z_{-}(\theta^*_{+})$ for type (iii) trajectory in the upper half of the channel and $z_0 = z_{-}(\theta^*_{-})$ otherwise. We call $B^*(\eta)$ the function of  eq.~(\ref{z_thetaB}) when choosing these initial conditions. We then derive for the (iii) and (iv) trajectories, explicit integral expressions for $\Delta x$ and  $\Delta y$ provided the expressions:
%
\begin{equation}
\begin{gathered}
\begin{aligned}
K_x &= (r^2+1)\sqrt{\frac{a^2-\eta^{*2}}{r^2-1}}\\
K_y &= \frac{\cos(\phi_0)}{| \cos(\phi_0)| } \frac{(1-\eta^{*2})(r^2+1)}{\sqrt{r^2-1}},
\end{aligned}
\end{gathered}
\end{equation}
-~Type (iii) trajectories -~ \cor{These} trajectories cross the planes $\phi=0$ or $\phi=\pi$. The integrals are split into two domains $[\eta^* , -\eta^*]$ and  $[-\eta^* , \eta^*]$ yielding two identical contributions, then:
\begin{equation}
\begin{gathered}
\begin{aligned}
\Delta x &= 2 K_x \int_{\eta^*}^{-\eta^*}\frac{(A\eta-B^*(\eta))d\eta}{(a^2-\eta^2)\sqrt{(\eta^{*2}-\eta^2)(1+4B^*(\eta))}}\\ 
\Delta y &= 2 K_y \int_{\eta^*}^{-\eta^*}\frac{Ad\eta}{\sqrt{(1+4B^*(\eta))(a^2-\eta^2)(\eta^{*2}-\eta^2)}}\\ 
\end{aligned}
\end{gathered}
\end{equation}\\
%

%
-~Type (iv) trajectories -~  the trajectories cross the planes $z=1/2$ and $\phi=\pi$. The integrals are split into 4 domains $[\eta^* , \eta_{1/2}]$, $[\eta_{1/2}, \eta^*]$ , $[\eta^* , \eta_{1/2}]$ and $[\eta_{1/2}, \eta^*]$, yielding four identical contributions, then:
\begin{equation}\label{Deltax}
\begin{gathered}
\begin{aligned}
\Delta x &=  4K_x\int_{\eta^*}^{\eta_{1/2}}\frac{(A\eta-B^*(\eta))d\eta}{(a^2-\eta^2)\sqrt{(\eta^{*2}-\eta^2)(1+4B^*(\eta))}}\\
\Delta y &=  4K_y\int_{\eta^*}^{\eta_{1/2}}\frac{Ad\eta}{\sqrt{(1+4B^*(\eta))(a^2-\eta^2)(\eta^{*2}-\eta^2)}}\\
\end{aligned}
\end{gathered}
\end{equation}

Then dividing $\Delta y$ by $\Delta x$ we obtain an analytical expression for $\tan\psi$:

\begin{equation}\label{TanPsiType3}
\tan\psi = \epsilon_y \frac{1-\eta^{*2}}{\sqrt{a-\eta^{*2}}} \frac{\int_{\eta^*}^{\eta_f}\frac{A d\eta}{\sqrt{(a^2-\eta^2)(\eta^{*2}-\eta^2)(1+4B^*(\eta))}}}{\int_{\eta^*}^{\eta_f}\frac{(A \eta -B^*(\eta)) d\eta}{(a^2-\eta^2)\sqrt{(1+4B*(\eta))(\eta^{*2}-\eta^2)}}}
\end{equation}
where $\epsilon_y = \frac{\cos(\phi_0)}{|\cos(\phi_0)| }$ is a sign function, positive for an initial swimmer orientation towards increasing $y$ (otherwise negative). The function $B^*(\eta)$ is defined by eq.~(\ref{z_thetaB}) \cor{with $\theta_0=\theta^*_{-}$ and $z_0 = z_{+}(\theta^*_{-})$} for type (iii) trajectories in the upper half of the channel and by $z_{-}(\theta^*_{-})$ otherwise. 
For type (iii) trajectories, $\eta_{f}=\cos (\theta_{1/2})$, and for type (iv), $\eta_{f}=\cos (\theta_{1/2})$.\\
To obtain $\tan\psi$, we compute the integrals with Matlab using the function {\it trapz}.

\subsection{Equations of the orientation with noise}

In the section ``Influence of thermal orientational noise'' the numerical trajectories of fig. (6) have been obtained by adding a noise of amplitude $D_r$ to the equation of the evolution of the orientation $\mathbf{p}$. 
Where $D_r$ is the thermal rotational diffusion coefficient of an ellipsoid of length $l$ and width $e$ immersed in a fluid of a viscosity $\eta$ at a temperature $T$. Its expression is given by the formula:

\begin{equation}\begin{gathered}
\begin{aligned}
D_r  &= \frac{3k_bT\ln({{2l}/{a}})}{\pi\eta l^3}
\end{aligned}
\end{gathered}
\end{equation}

The evolution of the orientation $\mathbf{p}$ is then:

\begin{equation}\label{kinematics_angles_noise} 
\begin{gathered}
\begin{aligned}
\dot{\mathbf{p}} &=(\overline{\overline{\mathbf{I}}}-\mathbf{p}\mathbf{p})(\beta \overline{\overline{\mathbf{E}}}+\overline{\overline{\mathbf{\Omega}}})\mathbf{p} +\sqrt{\frac{2}{{\rm Pe}}}~\mathbf{p}\land \pmb{\xi}
\end{aligned}
\end{gathered}
\end{equation}
where $\pmb{\xi}$ is a vectorial white noise with $\langle\xi_i(t)\rangle=0$ and $\langle\xi_i(t_1)\xi_j(t_2)\rangle=\delta(t_1-t_2)\delta_{ij}$,  ${\rm Pe }= \frac{\dot{\gamma}_M}{D_r}$ \cor{and the Stratonovich interpretation must be used for the multiplicative noise}.

The projections on the $x$, $y$ and $z$ axes are then :
\begin{equation}\label{kinematics_angles_noise} 
\begin{gathered}
\begin{aligned}
\dot{p_x} &= \frac{1}{2}p_z\big[\beta(1-2p_x^2)+1\big](1-2z) + \sqrt{\frac{2}{{\rm Pe}}}(p_z\xi_y-p_y\xi_z) \\
\dot{p_y} &= -\beta p_x p_y p_z(1-2z) + \sqrt{\frac{2}{Pe}}(-p_z\xi_x+p_x\xi_z) \\
\dot{p_z} &= \frac{1}{2}p_x\big[\beta(1-2p_z^2)+1\big](1-2z) + \sqrt{\frac{2}{{\rm Pe}}}(p_y\xi_x-p_x\xi_y)
\end{aligned}
\end{gathered}
\end{equation}
\cor{and the corresponding discretization scheme is:}

\begin{equation}\label{discretization_scheme}
\begin{split}
\begin{gathered}
\begin{aligned}
d{p_x} &= \frac{dt}{2}p_z\big[\beta(1-2p_x^2)+1\big](1-2z) \\
&+ \sqrt{\frac{2dt}{{\rm Pe}}}(p_z\xi_y-p_y\xi_z) - \frac{2dt}{P_e}p_x \\
d{p_y} &= -dt\beta p_x p_y p_z(1-2z)  \\
&+ \sqrt{\frac{2dt}{Pe}}(-p_z\xi_x+p_x\xi_z) - \frac{2dt}{P_e}p_y \\
d{p_z} &= \frac{dt}{2}p_x\big[\beta(1-2p_z^2)+1\big](1-2z)  \\
&+ \sqrt{\frac{2dt}{{\rm Pe}}}(p_y\xi_x-p_x\xi_y) - \frac{2dt}{P_e}p_z \\
\end{aligned}
\end{gathered}
\end{split}
\end{equation}
where the  additional term $-2dt\mathbf{p}/P_e$ corresponds to the drift produced by the multiplicative noise in \eqref{kinematics_angles_noise} \cite{gardiner1985handbook}. Finally, after each time step \eqref{discretization_scheme}, $\textbf{p}$ is rescaled to obtain a normalized vector.

\end{document}